\documentclass[a4paper,12pt,twoside]{article}
\usepackage{a4wide}

\usepackage{color,colordvi,graphicx,xcolor}
\usepackage[pdftex,colorlinks,bookmarksnumbered,bookmarksopen,citecolor=red,urlcolor=red]{hyperref}
\usepackage{amssymb,amsmath,amsthm}
\usepackage{yhmath} 
\usepackage{mathdots}
\usepackage[only,llbracket,rrbracket,interleave]{stmaryrd}

\usepackage{array}
\usepackage{multirow}
\usepackage{tabularx}
\usepackage{extarrows}
\usepackage{booktabs}

\usepackage{caption}
\usepackage{enumitem}

\newcommand{\bm}[1]{\text{\boldmath $#1$\unboldmath}}

\newcommand{\abs}[1]{\lvert#1\rvert}

\newcommand{\grad}{\boldsymbol{\nabla}}
\newcommand{\bu}{\bm{u}}

\newcommand{\bd}{\bm{d}}
\newcommand{\bx}{\bm{x}}
\newcommand{\bn}{\bm{n}}
\newcommand{\bs}{\bm{s}}
\newcommand{\bt}{\bm{t}}
\newcommand{\bnJ}{\bn_{j}}
\newcommand{\btJ}{\bt_{j}}
\newcommand{\buJ}{\bu_{j}}
\newcommand{\butJ}{\bm{\tilde{u}}_{j}}

\newcommand{\eltwo}{\ensuremath{\mathcal{L}^2}}
\newcommand{\Ga}[1]{\Gamma_{\!#1}}
\newcommand{\Aset}{\mathcal{A}_e}

\newcommand{\Lf}{L_{f}}
\newcommand{\Uo}{U_{o}}

\newcommand{\CellE}{\Omega_e}
\newcommand{\CellL}{\Omega_{\ell}}
\newcommand{\FaceJ}{\Gamma_{\!e,j}}
\newcommand{\areaFj}{\abs{\FaceJ}}
\newcommand{\buE}{\bu_{e}}
\newcommand{\buL}{\bu_{\ell}}
\newcommand{\bxE}{\bx_{e}}
\newcommand{\bxL}{\bx_{\ell}}
\newcommand{\bxJ}{\bx_{j}}
\newcommand{\bxtJ}{\bm{\tilde{x}_{j}}}

\newcommand{\gradUJ}{\grad\bu|_{j}}
\newcommand{\gradUmJ}[1]{\grad\bu^{#1}\rvert_{j}}

\newcommand{\bSf}{\bm{S}_{f}}
\newcommand{\FaceF}{\Gamma_{\!e,f}}
\newcommand{\areaFf}{\abs{\FaceF}}
\newcommand{\bnf}{\bn_{f}}
\newcommand{\btf}{\bt_{f}}
\newcommand{\bDelta}{\bm{\Delta}}
\newcommand{\bk}{\bm{k}}

\newcommand{\nsd}  {\ensuremath{\texttt{n}_{\texttt{sd}}}}
\newcommand{\numel}{\ensuremath{\texttt{n}_{\texttt{el}}}}
\newcommand{\numfa}{\ensuremath{\texttt{n}_{\texttt{fa}}}}
\newcommand{\numfael}{\ensuremath{\texttt{n}_{\texttt{fa}}^{\texttt{e}}}}

\newcommand{\etal}{\emph{et al.}\,}

\usepackage{scalerel,stackengine}
\stackMath
\newcommand\reallyhat[1]{%
\savestack{\tmpbox}{\stretchto{%
  \scaleto{%
    \scalerel*[\widthof{\ensuremath{#1}}]{\kern-.6pt\bigwedge\kern-.6pt}%
    {\rule[-\textheight/2]{1ex}{\textheight}}
  }{\textheight}%
}{0.5ex}}%
\stackon[1pt]{#1}{\tmpbox}%
}

\newenvironment{keywords}{\begin{quote}\emph{\textbf{Keywords:}}}{\end{quote}}

\theoremstyle{definition}

\newtheorem{remark}{Remark}

\usepackage{subfigure}
\usepackage{tikz}
\usetikzlibrary{fadings}
\usetikzlibrary{patterns}
\usetikzlibrary{shadows.blur}
\usetikzlibrary{shapes}
\usetikzlibrary{calc,intersections,arrows.meta,angles,quotes}

\graphicspath{ {figs/} }

\begin{document}
\title{An OpenFOAM face-centred solver for incompressible flows robust to mesh distortion}

\author{
\renewcommand{\thefootnote}{\arabic{footnote}}
			Davide Cortellessa\footnotemark[1] , \
			Matteo Giacomini\footnotemark[1]\textsuperscript{ \ ,}\footnotemark[2]\textsuperscript{ \ ,}*  ,\  and
			Antonio Huerta\footnotemark[1]\textsuperscript{ \ ,}\footnotemark[2]
}

\date{}
\maketitle

\renewcommand{\thefootnote}{\arabic{footnote}}

\footnotetext[1]{Laboratori de C\`alcul Num\`eric (LaC\`aN), ETS de Ingenier\'ia de Caminos, Canales y Puertos, Universitat Polit\`ecnica de Catalunya, Barcelona, Spain.}
\footnotetext[2]{Centre Internacional de M\`etodes Num\`erics en Enginyeria (CIMNE), Barcelona, Spain.
\vspace{5pt}\\
* Corresponding author: Matteo Giacomini \textit{E-mail:} \texttt{matteo.giacomini@upc.edu}
}

\begin{abstract}
This work presents an overview of mesh-induced errors commonly experienced by cell-centred finite volumes (CCFV), for which the face-centred finite volume (FCFV) paradigm offers competitive solutions.
In particular,  a robust FCFV solver for incompressible laminar flows is integrated in OpenFOAM and tested on a set of steady-state and transient benchmarks.
The method outperforms standard \texttt{simpleFoam} and \texttt{pimpleFoam} algorithms in terms of optimal convergence, accuracy, stability, and robustness.
Special attention is devoted to motivate and numerically demonstrate the ability of the FCFV method to treat non-orthogonal, stretched, and skewed meshes, where CCFV schemes exhibit shortcomings.
\end{abstract}


\section{Introduction}
\label{sc:Intro}

An appropriate design of computational meshes is crucial to achieve accurate, stable, and robust numerical solutions. 
Common mesh quality metrics include \emph{smoothness}, \emph{stretching}, \emph{orthogonality}, and \emph{skewness}. 
Smoothness accounts for mesh size variations between two contiguous cells and is particularly relevant in boundary layer regions where transition should not exceed $15 {\sim} 20\%$.
Stretching denotes cell aspect ratio, a critical mesh feature in the presence of large gradients of the solution.
Orthogonality is associated with the angular deviation of the vector connecting the centroids of two contiguous cells from the normal vector to their shared face.
Skewness measures the distance between the barycentre of a face and its intersection with the vector connecting the centroids of the neighbouring cells.
Whilst the first two aspects represent common criteria to be accounted for during mesh generation for many numerical methods (e.g., finite elements, spectral methods, finite volumes, \ldots), orthogonality and skewness introduce design constraints specific for finite volume (FV) schemes~\cite{barth2017}.

Indeed, despite their appealing efficiency and robustness, FV methods are known to be especially sensitive to mesh design.
The need to perform a reconstruction of the gradient of the solution at mesh faces is responsible for a loss of accuracy in the presence of distorted, stretched, and skewed meshes~\cite{diskin2010,diskin2011}.
The face-centred finite volume (FCFV) method, first proposed in~\cite{sevilla2018}, remedies this issue by devising a novel mixed hybrid FV fomulation circumventing the need for gradient reconstruction. This significantly reduces the sensitivity of the method to mesh-induced errors, while appealing properties of accuracy, robustness, and stability in the incompressible limit are inherited from the hybridisable discontinuous Galerkin framework~\cite{cockburn2009al}.

In this work,  the FCFV formulation for laminar incompressible Navier-Stokes flows presented in~\cite{vieira2024} is integrated in the open-source computational fluid dynamics (CFD) software OpenFOAM~\cite{jasak1996,openfoamUserGuide}. 
By significantly reducing the complexity of mesh generation, allowing for unstructured simplicial meshes to be seamlessly employed in OpenFOAM, the FCFV paradigm offers a competitive alternative to existing cell-centred finite volume (CCFV) routines such as \texttt{simpleFoam}, \texttt{pisoFoam}, and \texttt{pimpleFoam}.
Particular attention is devoted to discuss the sources of mesh-induced errors in CCFV, highlighting the concurrent robustness of the FCFV method in the presence of non-orthogonal, stretched, and skewed meshes.

The remainder of this chapter is organised as follows. 
Section~\ref{sc:Sensitivity} presents two common sources of mesh-induced errors in CCFV methods and explains the robustness of the FCFV paradigm in these scenarios. 
Section~\ref{sc:FCFV-OF} outlines the key novelties associated with the integration of FCFV in OpenFOAM. 
Finally, numerical examples are presented in Section~\ref{sc:Numerical}, and Section~\ref{sc:Conclusion} summarises the contributions of the work.

\section{Mesh-induced errors in cell-centred finite volumes}
\label{sc:Sensitivity}

This section focusses on two common sources of mesh-induced errors --mesh non-orthogonality and face skewness-- for OpenFOAM CCFV schemes, their effects on robustness and accuracy, and the robustness of the FCFV method. To simplify the presentation, but without loss of generality, the discussion is presented for 2D cases.

\subsection{Error induced by mesh non-orthogonality}
\label{sc:Orthogonality}

Consider two neighbouring cells $\CellE$ and $\CellL$, with centroids $\bxE$ and $\bxL$, as depicted in Figure~\ref{fig:NonOrth}. The vector $\bd {:=} \bxL{-}\bxE$ denotes the distance between the centroids of the cells, whereas $\bnJ$ and $\btJ$ are the unit normal and tangent vectors to the $j$-th face $\FaceJ$ of cell $\CellE$.
\begin{figure}[!htb]
\centering
\subfigure[Orthogonal cells ($\theta {=} 0$).]{
\begin{tikzpicture}[scale=1.4]
    \coordinate (A) at (0,0);
    \coordinate (B) at (2,-1.5);
    \coordinate (C) at (2,1);
    \coordinate (D) at (5,0);

    \draw [line width = 1, line join=round] (A) -- (B) -- (C)  node[above] {$\FaceJ$} -- node[midway, above left] {$\CellE$} cycle;
    \draw [line width = 1, line join=round] (C) -- (B) -- (D) -- node[midway, above right] {$\CellL$} cycle;
    
    \coordinate (C1) at (barycentric cs:A=1,B=1,C=1);
    \coordinate (C2) at (barycentric cs:C=1,B=1,D=1);    
  
    \draw[-{Latex[length=2mm]},line width = 1, color=blue,name path=line1] (C1)  -- (C2) node[above left] {$\bd$} ;

    \path [name path=lineBC] (B) -- (C);
    \path [name intersections={of=line1 and lineBC,by=Intersection}];
    
    \coordinate (PerpStart) at (Intersection);
    \coordinate (PerpEnd) at ($(Intersection)!0.5cm!90:(B)$);
    \coordinate (TangEnd) at ($(Intersection)!0.5cm!(C)$);
    
    \draw[-{Latex[length=2mm]},line width = 1,color=red] (PerpStart) -- (PerpEnd) node[midway,below] {$\bnJ$};
    \draw[-{Latex[length=2mm]},line width = 1,color=red] ([yshift=-0.02cm]PerpStart) -- (TangEnd) node[midway,left] {$\btJ$};
    
    \fill (C1) circle (1pt) node[left] {$\bxE$};
    \fill (C2) circle (1pt) node[right] {$\bxL$};

\end{tikzpicture}
\label{fig:NonOrthA}}
\qquad
\subfigure[Non-orthogonal cells.]{
\begin{tikzpicture}[scale=1.4]
    \coordinate (A) at (0,-1.5);
    \coordinate (B) at (2,-1.5);
    \coordinate (C) at (2,0.8);
    \coordinate (D) at (5,1.5);

    \draw [line width = 1, line join=round] (A) -- (B) -- (C) node[above] {$\FaceJ$} -- node[midway, above left] {$\CellE$} cycle;
    \draw [line width = 1,line join=round] (C) -- (B) -- node[midway, below right] {$\CellL$} (D) -- cycle;
    
    \coordinate (C1) at (barycentric cs:A=1,B=1,C=1);
    \coordinate (C2) at (barycentric cs:C=1,B=1,D=1);    
     
    \draw[-{Latex[length=2mm]},line width = 1, color=blue,name path=line1] (C1) -- (C2) node[above left] {$\bd$} ;

    \path [name path=lineBC] (B) -- (C);
    \path [name intersections={of=line1 and lineBC,by=Intersection}];
   
    \coordinate (PerpStart) at (Intersection);
    \coordinate (PerpEnd) at ($(Intersection)!0.5cm!90:(B)$);
    \coordinate (TangEnd) at ($(Intersection)!0.5cm!(C)$);

    \coordinate (ExtendedLine) at ($(C1)!1.2!(C2)$);
    \draw[draw=none, name path=extendedLine] (C1) -- (ExtendedLine);
    \path [name intersections={of=extendedLine and lineBC,by=ExtendedIntersection}];

    \pic[draw, "$\theta$", angle eccentricity=2.2, angle radius=0.35cm,line width = 1,color=purple] {angle = PerpEnd--Intersection--ExtendedLine};
    
    \draw[-{Latex[length=2mm]},line width = 1, color=blue,name path=line1] (C1) -- (C2);
    
    \draw[-{Latex[length=2mm]},line width = 1,color=red] (PerpStart) -- (PerpEnd) node[midway, below] {$\bnJ$};
    
    \fill (C1) circle (1pt) node[below] {$\bxE$};
    \fill (C2) circle (1pt) node[right] {$\bxL$};    
    
    \draw[-{Latex[length=2mm]},line width = 1,color=red] ([yshift=-0.02cm]PerpStart) -- (TangEnd) node[midway,left] {$\btJ$};
   
\end{tikzpicture}
\label{fig:NonOrthB}}
\caption{Angle $\theta$ between the vector $\bd$ connecting the centroids of cells $\CellE$ and $\CellL$ and the vector $\bnJ$ normal to face $\FaceJ$.}
\label{fig:NonOrth}
\end{figure}

It is straightforward to observe that $\bd$ can be expressed in terms of the orthonormal reference system $\{\bnJ,\btJ\}$ as 
\begin{equation}\label{eq:OrthogonalDecomp}
\bd = \abs{\bd}(\cos{\theta}\,\bnJ + \sin{\theta}\,\btJ),
\end{equation}
where $\theta$ is the \emph{non-orthogonality angle} measuring the angular deviation of $\bd$ with respect to $\bnJ$.
In the case of orthogonal meshes, $\theta {=} 0$ and $\bd$ reduces to $\abs{\bd}\bnJ$. Nonetheless, for non-orthogonal meshes, the tangential component of $\bd$ along face $\FaceJ$ does not vanish and its magnitude increases with $\theta$. Indeed,  OpenFOAM utility \texttt{checkMesh} provides a warning when $\theta {>} 7\pi/18$, that is, when the weight of the tangential component is approximately three times larger than the normal one,  suggesting to either improve mesh quality or to employ non-orthogonality correction strategies.

In order to understand the influence of the geometric property described above on the CCFV approximation, consider the incompressible Navier-Stokes equations for viscous laminar flows.  In this context, the kinematic viscosity $\nu {=} 1/Re$ is constant, where $Re$ is the Reynolds number.
Let $(\bu,p)$ denote the velocity-pressure pair.
Integrating the viscous term of the momentum equation on cell $\CellE$ and applying the divergence theorem, it follows
\begin{equation}\label{eq:viscousTerm}
    \int_{\CellE} \grad\cdot(\nu\grad\bu)\,d\Omega
    =
    \int_{\partial\CellE}(\nu\grad\bu)\bn\, d\Ga{}
    \simeq
    \nu \! \sum_{j\in\Aset} \areaFj \gradUJ \bnJ ,
\end{equation}
where $\Aset$ is the set of all faces of cell $\CellE$,  $\areaFj$ is the area of face $\FaceJ$, and $\gradUJ$ is the gradient of velocity at the barycentre of $\FaceJ$.  The last step in~\eqref{eq:viscousTerm} stems from integrating the viscous flux $(\nu\grad\bu)\bn$ on each face $\FaceJ$ using one quadrature point.
\\
The computation of~\eqref{eq:viscousTerm} requires the evaluation of the gradient of velocity in the normal direction to the face.  Under the assumption $\theta {<} \pi/2$,  $\bnJ$ can be rewritten in terms of $\bd$ and $\btJ$ by means of~\eqref{eq:OrthogonalDecomp}, thus leading to
\begin{equation}\label{eq:normalRewritten}
\bnJ = \frac{1}{\cos{\theta}}\frac{\bd}{\abs{\bd}} - \tan{\theta} \, \btJ .
\end{equation}
Inserting~\eqref{eq:normalRewritten} into~\eqref{eq:viscousTerm}, the computation of the viscous term yields
\begin{equation}\label{eq:viscousTermSplit}
    \nu \! \sum_{j\in\Aset} \areaFj \gradUJ \bnJ
    =
    \nu \! \sum_{j\in\Aset} \areaFj \left(\frac{1}{\cos{\theta}} \gradUJ \frac{\bd}{\abs{\bd}} - \tan{\theta} \, \gradUJ \btJ \right) .
\end{equation}
On the one hand, the projection of the gradient of velocity along the unit vector $\bd/\abs{\bd}$ is computed by means of a second-order central difference scheme using the values $\buE$ and $\buL$ of the velocity at the centroid $\bxE$ and $\bxL$ of $\CellE$ and $\CellL$, respectively. 
On the other hand, the component of the gradient of velocity tangential to $\FaceJ$ is unknown and is employed by OpenFOAM to introduce the so-called \emph{non-orthogonal correction} by evaluating $\gradUJ \btJ$ using the gradient of the last computed velocity $\bu^{m-1}$. 
Hence, at iteration $m$ of the non-orthogonal correction,  the terms in~\eqref{eq:viscousTermSplit} are approximated as
\begin{equation}\label{eq:viscousTermCorr}
    \nu \! \sum_{j\in\Aset} \areaFj \gradUmJ{m} \bnJ
    =
    \nu \! \sum_{j\in\Aset} \areaFj \left(\frac{1}{\cos{\theta}} \frac{\buL^m-\buE^m}{\abs{\bd}} - \tan{\theta} \, \gradUmJ{m-1} \btJ \right) .
\end{equation}

Note that, for orthogonal meshes, the orientations of $\bd$ and $\bnJ$ coincide and $\theta {=}0$, see Figure~\ref{fig:NonOrthA}.  In this case,  the last term in~\eqref{eq:viscousTermCorr} vanishes and the computation of the normal flux across $\FaceJ$ reduces to the central difference scheme applied to the velocities $\buE$ and $\buL$ at the centroids $\bxE$ and $\bxL$.
On the contrary, for meshes with high degrees of non-orthogonality (i.e., $\theta {>} 7\pi/18$), the magnitude of the second term significantly increases and its explicit treatment negatively affects the stability of CCFV scheme, possibily leading to the divergence of the solution. To remedy this issue, numerical strategies to bound the non-orthogonal correction are available, although they are known to negatively affect the accuracy of the computed solution~\cite{openfoamUserGuide}.

\begin{remark}
Equation~\eqref{eq:viscousTermCorr} can be easily rewritten using standard OpenFOAM notation.
Note that OpenFOAM treats each component of the velocity field independently and denotes such a generic scalar variable as $\phi$.
Moreover, replace the indices $e$ and $\ell$ of cells $\CellE$ and $\CellL$ by $N$ and $P$, respectively,  and denote by $f$ the index $j$ of face $\FaceJ$. 
The so-called \emph{surface normal vector} is thus given by $\bSf {:=} \areaFf \bnf$ and is decomposed as $\bSf {=} \bDelta {+} \bk$, with $\bDelta$ being oriented along the direction of vector $\bd/\abs{\bd}$. 
Following~\cite{jasak1996},  the magnitude of $\bDelta$ is selected such that $\abs{\bDelta} {:=} \areaFf / \cos{\theta}$ and the resulting non-orthogonal direction is given by $\bk {:=} {-} \areaFf \tan{\theta} \, \btf$.
Hence, for a generic component $\phi$ of the velocity vector, equation~\eqref{eq:viscousTermCorr} can be recast as
\begin{equation}\label{eq:viscousTermOF}
    \nu \! \sum_{f} \grad\phi^m|_f \cdot \bSf
    =
    \nu \! \sum_{f} \left( \abs{\bDelta} \frac{\phi_N^m-\phi_P^m}{\abs{\bd}} + \grad\phi^{m-1}|_f \cdot \bk  \right) .
\end{equation}
\end{remark}

\subsubsection*{Robustness of face-centred finite volumes to mesh non-orthogonality}

The FCFV method relies on a mixed hybrid FV formulation featuring velocity, pressure and gradient of velocity as unknowns in the cell $\CellE$, as well as a \emph{hybrid} velocity on each cell face $\FaceJ$~\cite{sevilla2018,vieira2024}.
As such,  the FCFV framework does not require the gradient reconstruction strategy described in~\eqref{eq:viscousTermCorr} to compute the inter-cell flux.
Indeed, the introduction of the so-called \emph{mixed variable} representing a scaling of the gradient of velocity and the definition of the inter-cell numerical flux in terms of the hybrid velocity allow to circumvent the need to access information of the neighbouring cell $\CellL$, thus avoiding any possible issue due to the angular deviation of $\bd$ with respect to $\bnJ$.

\subsection{Error induced by mesh skewness}
\label{sc:Skewness}

The second source of mesh-induced errors in CCFV schemes is face skewness, also known as \emph{mesh non-conjunctionality}~\cite{hill2018}.
This geometric feature is associated with the deviation of the barycentre $\bxJ$ of $\FaceJ$ from the intersection $\bxtJ$ of the vector $\bd$ connecting the centroids of two neighbouring cells and the face itself, as reported in Figure~\ref{fig:Skew}.
Let $\bs {:=} \bxJ {-} \bxtJ$ be the distance introduced above.  OpenFOAM utility \texttt{checkMesh} evaluates the face skewness by computing $\abs{\bs}/\abs{\bd}$ for each cell face.
\begin{figure}[!htb]
\centering
\subfigure[Non-skewed cells ($\bs {=} \bm{0}$).]{
\begin{tikzpicture}[scale=1.6]
    \coordinate (A) at (0,-2);
    \coordinate (B) at (1.4,-2);
    \coordinate (C) at (1.4,0);
    \coordinate (D) at (3,-2);
    \coordinate (E) at (3,0);
    \coordinate (F) at (0,0);

    \draw [line width = 1, line join=round] (A) --  (B) node[below] {$\FaceJ$} -- (C) -- node[midway, above left] {$\CellE$} (F) -- (A);
    \draw [line width = 1, line join=round] (C) -- (B) -- (D) -- (E) -- node[midway, above right] {$\CellL$} (C);

    \coordinate (C1) at (barycentric cs:A=1,B=1,C=1,F=1);
    \coordinate (C2) at (barycentric cs:C=1,B=1,D=1,E=1);
        
    \draw[-{Latex[length=2mm]},line width = 1, color=blue,name path=line1] (C1) -- (C2) node[above left] {$\bd$} ;
    
    \path [name path=lineBC] (B) -- (C);
    \path [name intersections={of=line1 and lineBC,by=Intersection}];
    
    \coordinate (PerpStart) at (Intersection);
    \coordinate (PerpEnd) at ($(Intersection)!0.5cm!90:(B)$);
    
    \draw[-{Latex[length=2mm]},line width = 1,color=red] (PerpStart) -- (PerpEnd) node[midway, below] {$\bnJ$};

    \coordinate (MidBC) at ($(B)!0.5!(C)$);
    \path [name path=lineBC] (B) -- (C);
    \path [name intersections={of=line1 and lineBC,by=F}];
        
    \fill (C1) circle (1pt) node[left] {$\bxE$};
    \fill (C2) circle (1pt) node[right] {$\bxL$};
    
    \draw [line width = 1] (MidBC) circle (1pt);
    \fill[white] (MidBC) circle (1pt);
    \node[above left] at (MidBC) {$\bxtJ {\equiv} \bxJ$};

\end{tikzpicture}
\label{fig:SkewA}}
\qquad
\subfigure[Skewed cells.]{
\begin{tikzpicture}[scale=1.6]
    \coordinate (A) at (1,-2.8);
    \coordinate (B) at (2,-2);
    \coordinate (C) at (2,0);
    \coordinate (D) at (4,-2.5);
    \coordinate (E) at (4.5,-1.9);
    \coordinate (F) at (0,-1.6);

    \draw [line width = 1, line join=round] (A) --  (B) -- (C) -- node[midway, above left] {$\CellE$} (F) -- (A);
    \draw [line width = 1, line join=round] (C) -- (B) node[below,xshift=1ex] {$\FaceJ$} -- (D) -- (E) -- node[midway, above right] {$\CellL$} (C);

    \coordinate (C1) at (barycentric cs:A=1,B=1,C=1,F=1);
    \coordinate (C2) at (barycentric cs:C=1,B=1,D=1,E=1);
        
    \draw[-{Latex[length=2mm]},line width = 1, color=blue,name path=line1] (C1) -- (C2) node[below left] {$\bd$} ;
    
    \path [name path=lineBC] (B) -- (C);
    \path [name intersections={of=line1 and lineBC,by=Intersection}];
    
    \coordinate (PerpStart) at (Intersection);
    \coordinate (PerpEnd) at ($(Intersection)!0.5cm!90:(B)$);
    
    \draw[-{Latex[length=2mm]},line width = 1,color=red] (PerpStart) -- (PerpEnd) node[midway, below] {$\bnJ$};

    \coordinate (MidBC) at ($(B)!0.5!(C)$);
    \path [name path=lineBC] (B) -- (C);
    \path [name intersections={of=line1 and lineBC,by=F}];
        
	\draw[-{Latex[length=2mm]},line width = 1, color=purple,name path=line1] (F) -- node[midway, right] {$\bs$} (MidBC) ;
    
    \fill (C1) circle (1pt) node[left] {$\bxE$};
    \fill (C2) circle (1pt) node[right] {$\bxL$};
    
    \draw [line width = 1] (MidBC) circle (1pt);
    \fill[white] (MidBC) circle (1pt);
    \node[left] at (MidBC) {$\bxJ$};

    \draw [line width = 1]  (F) circle (1pt);
    \fill[white] (F) circle (1pt);
    \node[below left] at (F) {$\bxtJ$};

\end{tikzpicture}
\label{fig:SkewB}}
\caption{Distance $\bs$ between the barycentre $\bxJ$ of face $\FaceJ$ and its intersection $\bxtJ$ with vector $\bd$ connecting the centroids of cells $\CellE$ and $\CellL$.}
\label{fig:Skew}
\end{figure}

OpenFOAM CCFV schemes require computational meshes with low and moderate skewness in order to preserve interpolation accuracy.
Consider the convective term in the momentum equation for an incompressible Navier-Stokes flow. Upon integration on cell $\CellE$ and application of the divergence theorem, it holds
\begin{equation}\label{eq:convectiveTerm}
    \int_{\CellE} \grad\cdot(\bu \otimes \bu)\,d\Omega
    =
    \int_{\partial\CellE}(\bu\cdot\bn)\bu\, d\Ga{}
    \simeq
    \sum_{j\in\Aset} \areaFj (\buJ \cdot \bnJ)\buJ ,
\end{equation}
where the last step follows from integrating the convective flux $(\bu\cdot\bn)\bu$ on each face $\FaceJ$ using one quadrature point.

It is worth noticing that equation~\eqref{eq:convectiveTerm} requires the velocity to be evaluated at the barycentre of the face. Nonetheless, face velocities are unknown in the CCFV framework and need to be computed from the values at the centroids of the two cells sharing the face under analysis.
This operation is performed in OpenFOAM by means of a linear interpolation,  leading to the computation of the velocity $\butJ$ at point $\bxtJ$ as a function of the velocities $\buE$ and $\buL$.

Note that, for non-skewed meshes,  the intersection $\bxtJ$ of $\bd$ and $\FaceJ$ coincides with the barycentre $\bxJ$ of the face and $\bs {=} \bm{0}$ (see Figure~\ref{fig:SkewA}).  In this case, the centred linear interpolation scheme automatically provides the value $\buJ$ of the velocity required to compute~\eqref{eq:convectiveTerm}.
On the contrary, when the meshes feature face skewness, the value of $\buJ$ could significantly differ from $\butJ$. This is indeed the case in the vicinity of physical walls, where the gradient of velocity is relevant. Hence, mesh skewness can negatively affect the accuracy of the solution in regions where physical walls are not planar.
In this context, OpenFOAM CCFV schemes determine the velocity at the barycentre of the face as
\begin{equation}\label{eq:skewCorr}
  \buJ = \butJ + \grad\bu|_{\bxtJ} \cdot \bs,
\end{equation}
where the second term, known as a \emph{skewness correction},  stems from a first-order Taylor expansion~\cite{openfoamUserGuide}. 
Nonetheless, it is worth highlighting that the resulting corrected interpolation strategy~\eqref{eq:skewCorr} is no longer second-order accurate, thus affecting the quality of the computed solution.

\subsubsection*{Robustness of face-centred finite volumes to mesh skewness}

As previously mentioned,  the FCFV method defines a hybrid variable at the barycentre of each face, thus circumventing the need to interpolate the velocity of neighbouring cells.
Moreover, as described in detail in~\cite{sevilla2018,vieira2024}, the FCFV paradigm expresses the velocity in each cell as a function of the face velocities by means of a \emph{hybridisation} procedure and the inter-cell communication of information is handled by enforcing the continuity of the normal fluxes in a weak sense.
This makes the method robust to the geometric features of the mesh, being able to treat a variety of cells (triangles and quadrilaterals in 2D, tetrahedra, hexahedra, prisms, and pyramids in 3D)~\cite{sevilla2018},  including non-planar quadrilaterals and skewed faces, as well as hybrid meshes combining different cell types~\cite{giacomini2020}.

\section{A paradigm for face-centred finite volumes in OpenFOAM}
\label{sc:FCFV-OF}

This section presents the main computational novelties related to the integration of a FCFV solver for incompressible flows in OpenFOAM.

The first and major novelty is represented by the hybrid nature of the FCFV scheme. This leads to a set of unknowns (velocity, pressure,  and gradient of velocity) being approximated as constant functions at the centroids of the cells and to a new,  piecewise constant face variable, known as hybrid velocity, being introduced at the barycentres of the mesh faces.
The degrees of freedom (DOFs) associated with velocity and gradient of velocity are expressed in terms of the face velocity and the cell pressure, by means of a hybridisation step.
The resulting FCFV global problem features a saddle-point structure, as typical in incompressible flow problems~\cite{Donea-Huerta-2003}. 
A detailed derivation of the formulation is available in~\cite{sevilla2018} for Stokes flow, and in~\cite{vieira2024} for laminar and turbulent incompressible Navier-Stokes, whereas readers interested in the FCFV method for inviscid and viscous compressible flows are referred to~\cite{vila2022}.

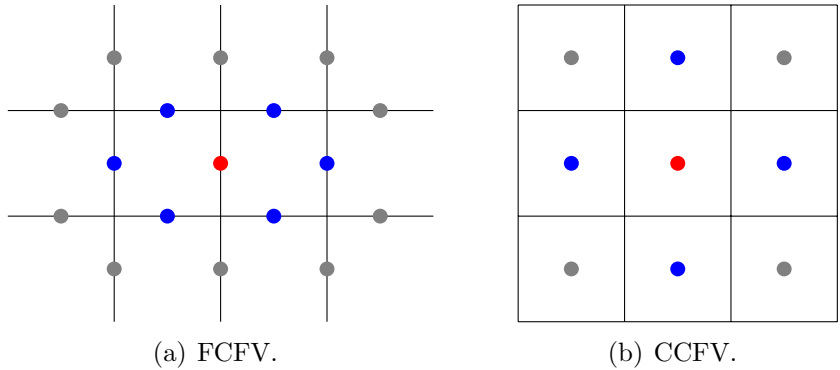
\begin{figure}[!htb]
\centering
\subfigure[FCFV.]{
\begin{tikzpicture}[scale=1.4]
    \draw[] (0,1) -- (4,1);
    \draw[] (0,2) -- (4,2);
    \draw[] (1,0) -- (1,3);
    \draw[] (2,0) -- (2,3);
    \draw[] (3,0) -- (3,3);

    \fill[gray] (1,0.5) circle (2pt);
    \fill[gray] (2,0.5) circle (2pt);
    \fill[gray] (3,0.5) circle (2pt);
    \fill[blue] (1,1.5) circle (2pt);
    \fill[red] (2,1.5) circle (2pt);
    \fill[blue] (3,1.5) circle (2pt);
    \fill[gray] (1,2.5) circle (2pt);
    \fill[gray] (2,2.5) circle (2pt);
    \fill[gray] (3,2.5) circle (2pt);
    \fill[gray] (0.5,1) circle (2pt);
    \fill[blue] (1.5,1) circle (2pt);
    \fill[blue] (2.5,1) circle (2pt);
    \fill[gray] (3.5,1) circle (2pt);
    \fill[gray] (0.5,2) circle (2pt);
    \fill[blue] (1.5,2) circle (2pt);
    \fill[blue] (2.5,2) circle (2pt);
    \fill[gray] (3.5,2) circle (2pt);
\end{tikzpicture}}
\qquad
\subfigure[CCFV.]{
\begin{tikzpicture}[scale=1.4]
    \draw[step=1cm] (0,0) grid (3,3);
    
    \fill[gray] (0.5,0.5) circle (2pt);
    \fill[gray] (0.5,2.5) circle (2pt);
    \fill[gray] (2.5,0.5) circle (2pt);
    \fill[gray] (2.5,2.5) circle (2pt);
    \fill[red] (1.5,1.5) circle (2pt);
    \fill[blue] (1.5,0.5) circle (2pt);
    \fill[blue] (1.5,2.5) circle (2pt);
    \fill[blue] (0.5,1.5) circle (2pt);
    \fill[blue] (2.5,1.5) circle (2pt);
\end{tikzpicture}}
\caption{Computational stencil of the discretisations on a mesh of quadrilateral cells. Red: node under analysis. Blue: nodes employed in the discretisation. Gray: inactive nodes.}
\label{fig:ConnectQua}
\end{figure}

The novel DOFs location has a direct consequence on the stencil employed for computation, as diplayed in Figure~\ref{fig:ConnectQua}.

Consider a mesh featuring only one type of cells, with $\numfael$ denoting the number of faces of cell $\CellE$.
In the FCFV discretisation, the unknown under analysis (i.e., the red node) is globally coupled to the DOFs located at the barycentres of all the faces of the two neighbouring cells.
On the contrary, in the CCFV scheme, each velocity unknown interacts with all the DOFs at the centroids of the neighbouring cells.
Hence,  both methods rely exclusively on first-neighbour connections and each DOF of the FCFV and CCFV schemes is coupled to $2(\numfael {-}1)$ and $\numfael$ unknowns, respectively.
Nonetheless, it is worth noticing that, on meshes without hanging nodes, the FCFV method only requires information of two cells per each face, whereas the CCFV scheme needs every cell to communicate with its $\numfael$ neighbours.
Moreover, contrary to the CCFV approach, the FCFV solver does not need to access the cell neighbours to perform interpolation on the faces, as such an information is naturally stored in the hybrid variable.

Of course, this increased flexibility entails an additional cost in terms of  problem size and storage requirements as the FCFV paradigm features more unknowns than the CCFV scheme.
Table~\ref{tab:DOFs} reports the estimated number of degrees of freedom for FCFV and CCFV approximations of incompressible flows, neglecting boundary unknowns.
\begin{table}[!htb]
    \centering
    \begin{tabular}{p{3cm} c c c c c}
        \hline
        Cell type    & Vertices & Cells & Faces & FCFV DOFs & CCFV DOFs \\
       \hline
        Quadrilaterals  & n     & n     & 2n    & 5n    & 3n \\
        Triangles       & n     & 2n    & 3n    & 8n    & 6n \\
        Tetrahedra      & n     & 5n    & 10n   & 35n   & 20n \\
        Hexahedra       & n     & n     & 3n    & 10n    & 4n \\
        Prisms          & n     & 2n    & 5n  & 17n    & 8n \\
        Pyramids        & n     & 8n/5  & 4n    & 68n/5 & 32n/5 \\
        \hline
    \end{tabular}
    \caption{Estimated number of degrees of freedom for FCFV and CCFV approximations.}    
    \label{tab:DOFs}
\end{table}
Moreover, the sparsity pattern of the FCFV global problem significantly differs from the standard CCFV one. 
It follows that the standard OpenFOAM \texttt{fvMatrix} class for sparse matrices is not suitable for a FCFV approximation, as it employs a fixed sparsity pattern based on the stencil used by cell-centred discretisations.
Hence, the OpenFOAM implementation of the FCFV method relies on the \texttt{lduMatrix} class, which inherits the core structure of OpenFOAM sparse matrices, but allows for user-defined sparsity patterns. 
The resulting sparse matrix is thus stored in three vectors, containing lower triangular, upper triangular, and diagonal components of the matrix.
From a practical viewpoint,  this information is encoded in a newly defined matrix addressing, namely the \texttt{fcfvAddressing}, which leverages OpenFOAM utilities and data structures. 
The entries on the diagonal are stored in an array of length $\nsd\numfa{+}\numel$, with $\numfa$ being the number of mesh faces where the $\nsd$ components of the velocity vector are defined and $\numel$ the number of cells where the pressure unknowns are located.
The entries of the lower and upper triangular portions of the matrix are stored using a standard sparse matrix structure featuring a total of four arrays: two arrays store the entries of the lower and upper triangular matrices, whereas arrays \textbf{\texttt{l}} and \textbf{\texttt{u}} contain the row and column indices of the non-zero entries of the upper triangular matrix, following a row-wise pattern.
It is worth noticing that although the FCFV global matrix is not symmetric for the Navier-Stokes problem, its sparsity pattern is, and $\textbf{\texttt{l}}^T$ and $\textbf{\texttt{u}}^T$ identify the indices of the non-zero entries of the lower triangular matrix.

\section{Numerical results}
\label{sc:Numerical}

Three numerical benchmarks are presented to evaluate the OpenFOAM FCFV solver for steady-state and transient incompressible laminar flows at different Reynolds numbers, with particular attention to its robustness to mesh-induced errors affecting accuracy, stability, and robustness of OpenFOAM CCFV schemes.
A more detailed discussion of the presented setups and results is available in~\cite{cortellessa2024}.

The FCFV numerical results are obtained using a Picard scheme for nonlinear iterations, with a tolerance of $10^{-6}$ on the relative residuals of the problem. For the transient problem in Section~\ref{sc:Oscillator},  this tolerance is imposed in the nonlinear solver at each time step,  allowing a maximum of five iterations. The FCFV global system is solved using a direct LU solver from the \texttt{MUMPS} library, available through \texttt{PETSc}~\cite{openfoamUserGuide}. 

For CCFV,  \texttt{simpleFoam} and \texttt{pimpleFoam} are employed for steady-state and transient cases, respectively. Second-order discretisation schemes are used in both space and time with non-orthogonal correction, setting  \texttt{nNonOrthogonalCorrectors} to 2.  To compute velocity prediction, the \texttt{PBICGStab} solver is used with \texttt{DILU} preconditioning, whereas the pressure equation is solved with the \texttt{PCG} solver and \texttt{DIC} preconditioning, both with a stopping criterion of $10^{-8}$. Under-relaxation is used in \texttt{simpleFoam}, with relaxation parameters $0.7$ and $0.3$ for velocity and pressure, respectively, and a tolerance of $10^{-6}$ for the nonlinear solver. For the transient case, \texttt{pimpleFoam} is employed with parameters \texttt{nCorrectors} and \texttt{nOuterCorrectors} equal to 3, and adaptive time step control with a maximum Courant number of $0.8$.

\subsection{Coaxial Couette flow}
\label{sc:Couette}

The coaxial Couette flow is defined on a circular annulus, with inner radius $R_i {=} 1$ and outer radius $R_o {=} 2$, where the angular velocities  $\omega_i {=} 0$ and  $\omega_o {=} 0.5$ are imposed.
This problem features an analytical solution independent of viscosity and the Reynolds number is set to $Re {=} 1$. For a detailed problem statement,  see~\cite{vieira2024}.

\begin{figure}[!htb]
	\centering
	\subfigure[Regular meshes.]{\includegraphics[width=0.3\textwidth]{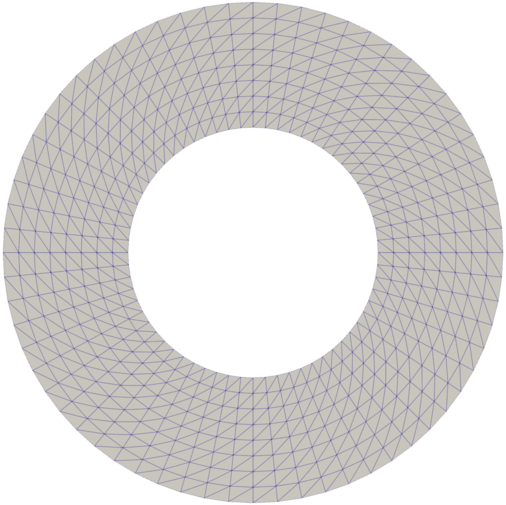}}
        \quad
	\subfigure[Distorted meshes.]{\includegraphics[width=0.3\textwidth]{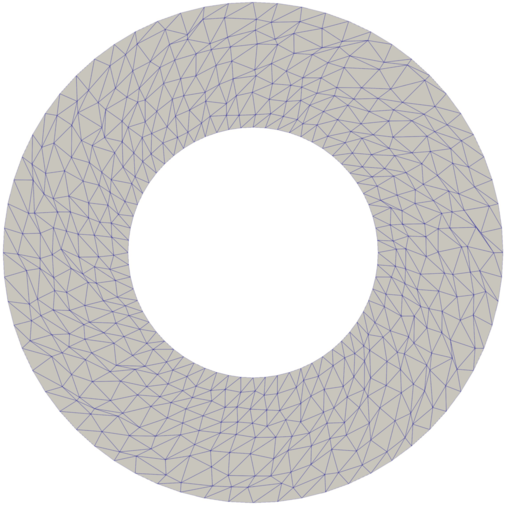}}
	\caption{Second level of refinement of the computational meshes for the Couette flow.}
	\label{fig:CouMesh}
\end{figure}

This problem is employed to assess the convergence properties of OpenFOAM FCFV and CCFV schemes, for a set of uniformly refined triangular meshes such that the $i$-th mesh has $(16{\times}2^i){\times}(2{\times}2^i){\times}2$ cells.  More precisely,  Figure~\ref{fig:CouMesh} reports the second level of refinement of a mesh of regular cells (left) and distorted cells (right), where the internal mesh nodes have been randomly displaced by up to $30\%$ of the smallest edge length emanating from a given node.

\begin{figure}[!htb]
	\centering
	\subfigure[FCFV.]{\includegraphics[width=0.45\textwidth]{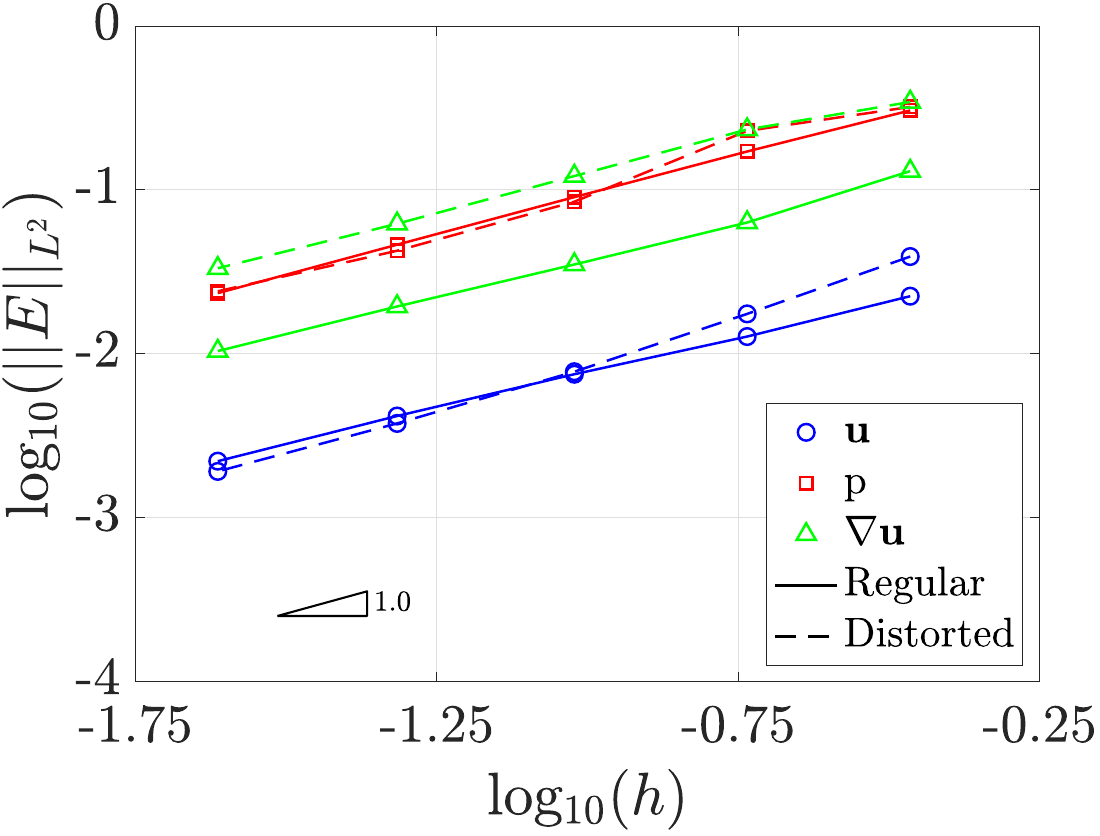}}
	\qquad
	\subfigure[CCFV.]{\includegraphics[width=0.45\textwidth]{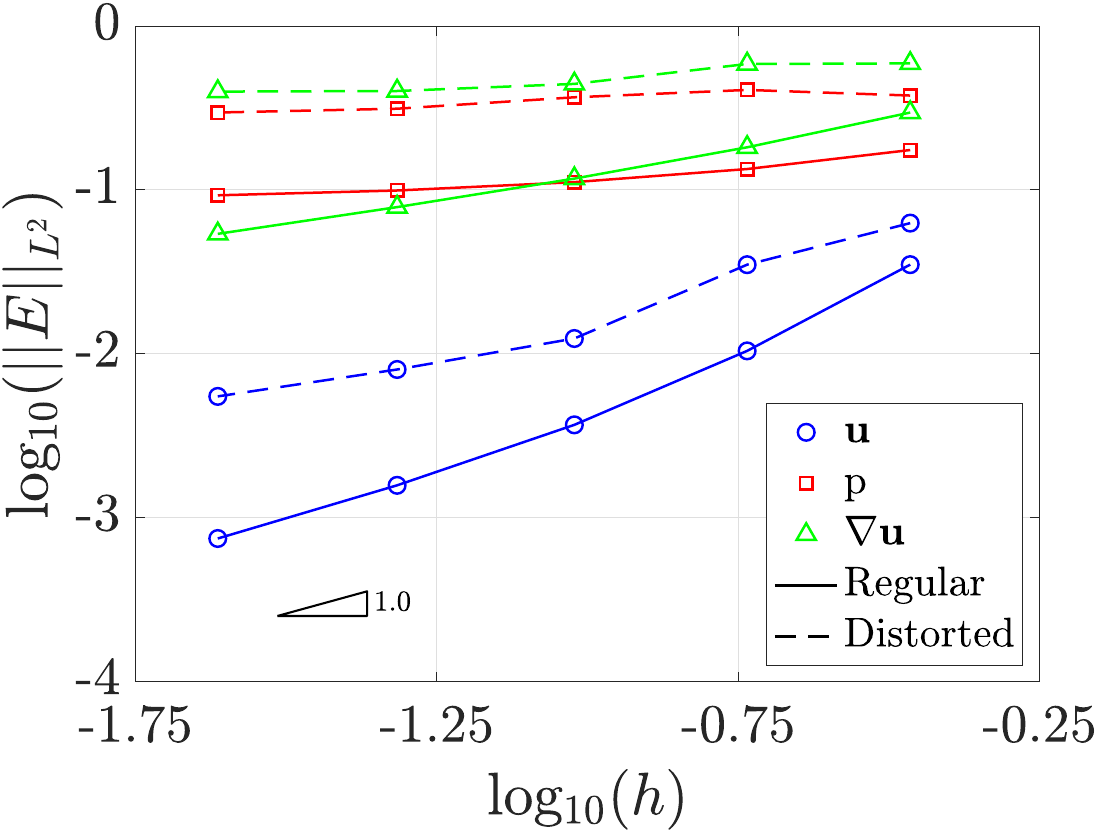}}
	\caption{Mesh convergence of the $\eltwo$ errors for the Couette flow.}
	\label{fig:CouTRI}
\end{figure}

Figure~\ref{fig:CouTRI} displays the $\eltwo$ norm of the relative error for velocity, pressure, and gradient of velocity as a function of the characteristic mesh size $h$. 
The OpenFOAM implementation of the FCFV method provides optimal convergence of order one for velocity, pressure, and gradient of velocity. More importantly, optimal convergence is preserved also when distorted meshes are employed, with accuracy perfectly matching for velocity and pressure, while slightly larger errors are obtained for the gradient of velocity.
On the contrary,  whilst it is well known that CCFV schemes achieve second-order convergence on structured, orthogonal, quadrilateral meshes, see, e.g., \cite{vila2023},  the results clearly show the inability of \texttt{simpleFoam} to converge on meshes of triangular cells. Indeed, while the method can predict reasonably well velocity, the error for pressure and gradient of velocity does not converge and stagnates around $10^{-1}$.  Moreover, these results significantly degrade when distorted grids are employed, with \texttt{simpleFoam} failing to reduce the error when $h$ decreases.
Hence, the OpenFOAM FCFV solver outperforms the CCFV approach and does not experience any numerical issue due to mesh non-orthogonality or skewness.
These results confirm the superior robustness to mesh distortion of the FCFV method with respect to CCFV schemes, previously observed for inviscid and viscous compressible flows (also in the incompressible limit)  in the comparison with Ansys Fluent solvers~\cite{vila2023}.

\subsection{Lid-driven cavity flow}
\label{sc:Cavity}

To evaluate the capability of the OpenFOAM FCFV scheme to treat convection phenomena, the lid-driven cavity in $\Omega {=} [0,1]^2$ is studied for $Re {=} 100$ and $Re {=} 1,000$.  The viscosity is set to $\nu {=} 1/Re$, a constant unit horizontal velocity is imposed on the top lid, and no-slip conditions are enforced on the remaining boundaries. 

\begin{figure}[!htb]
	\centering
	\subfigure[Regular meshes.]{\includegraphics[width=0.3\textwidth]{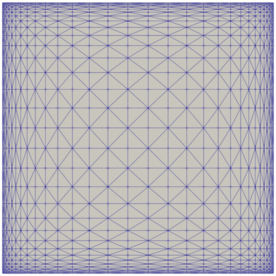}}
	\quad
        \subfigure[Distorted meshes.]{\includegraphics[width=0.3\textwidth]{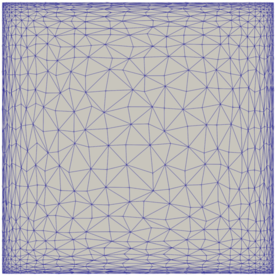}}
	\caption{First level of refinement of the computational meshes for the cavity flow.}
	\label{fig:CavMesh}
\end{figure}

For the mesh convergence study,  two sets of meshes (regular and distorted, see Figure~\ref{fig:CavMesh}) are considered, with local refinement near the physical walls. 
Besides mesh distortion, this case also assesses the suitability of the FCFV method to handle stretched cells,  with a maximum aspect ratio of approximately 30.
Mesh nodes are clustered near the boundaries using a hyperbolic tangent profile, see~\cite{Donea-Huerta-2003}.
The $i$-th computational mesh features $(16 {\times} 2^i) {\times} (16 {\times} 2^i) {\times} 2$ cells. 

\begin{figure}[!htb]
	\centering
	\subfigure[FCFV,  $Re {=} 100$.]{\includegraphics[width=0.45\textwidth]{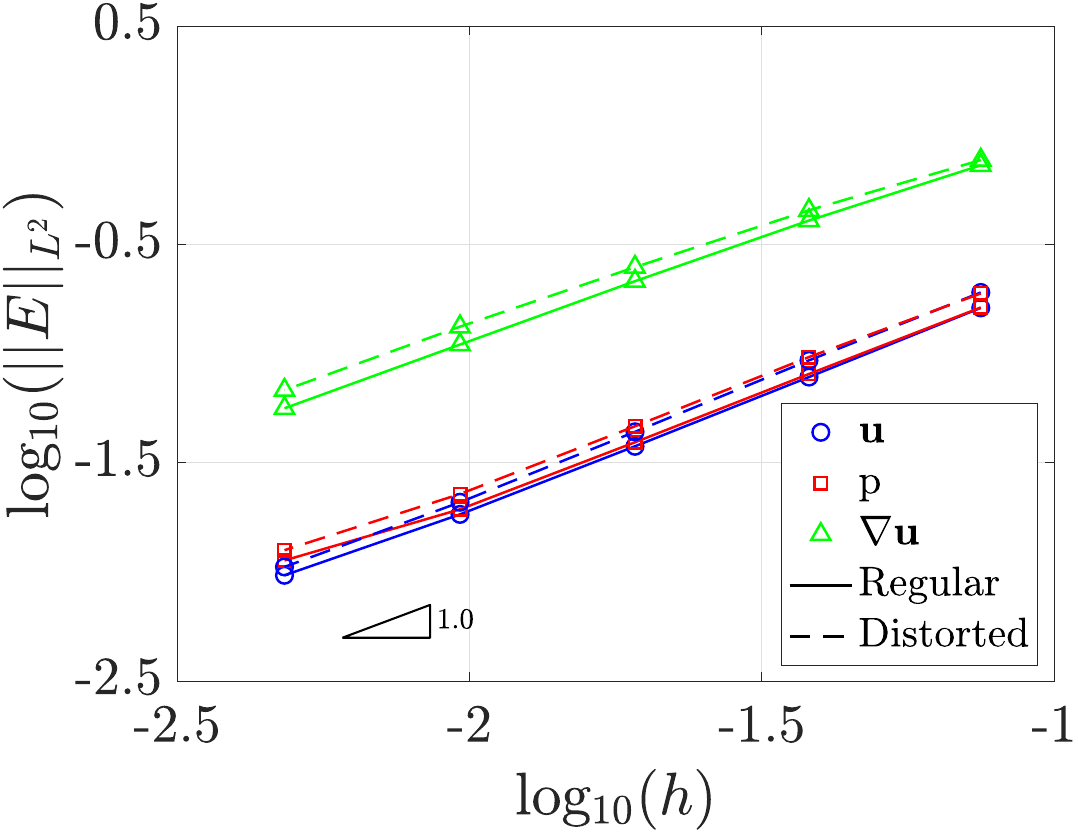}}
	\qquad
	\subfigure[CCFV, $Re {=} 100$.]{\includegraphics[width=0.45\textwidth]{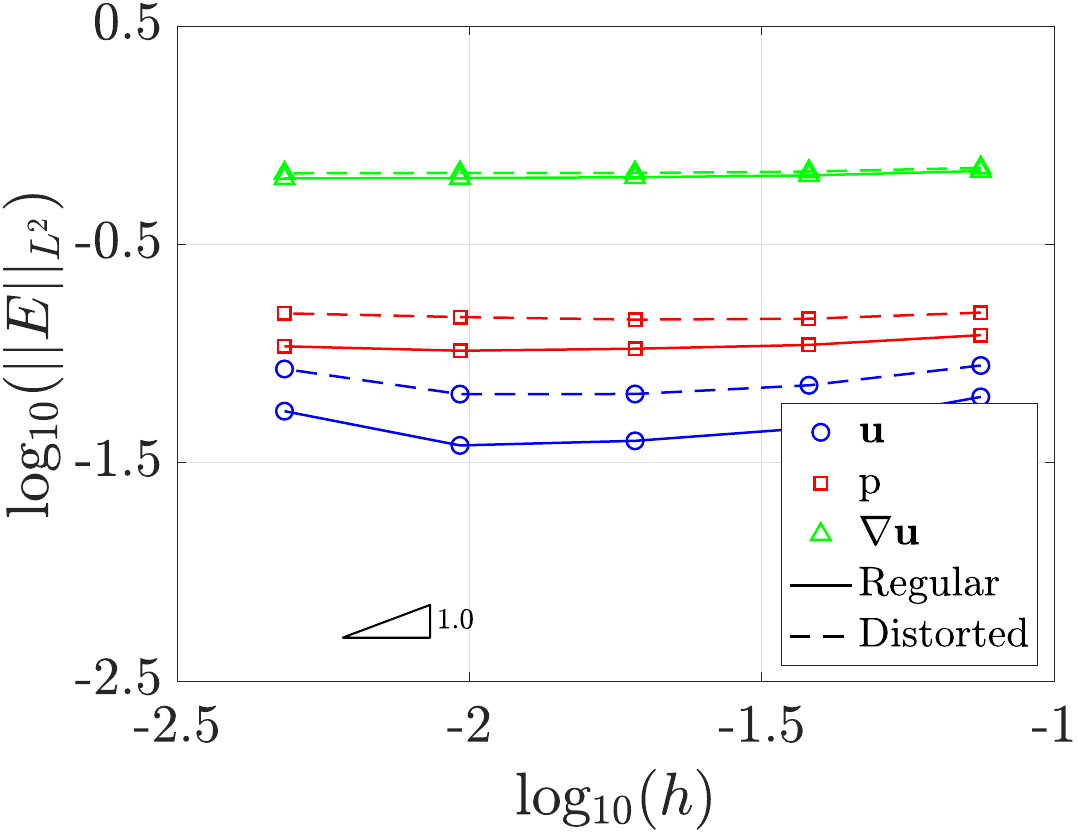}}

	\subfigure[FCFV, $Re {=} 1,000$.]{\includegraphics[width=0.45\textwidth]{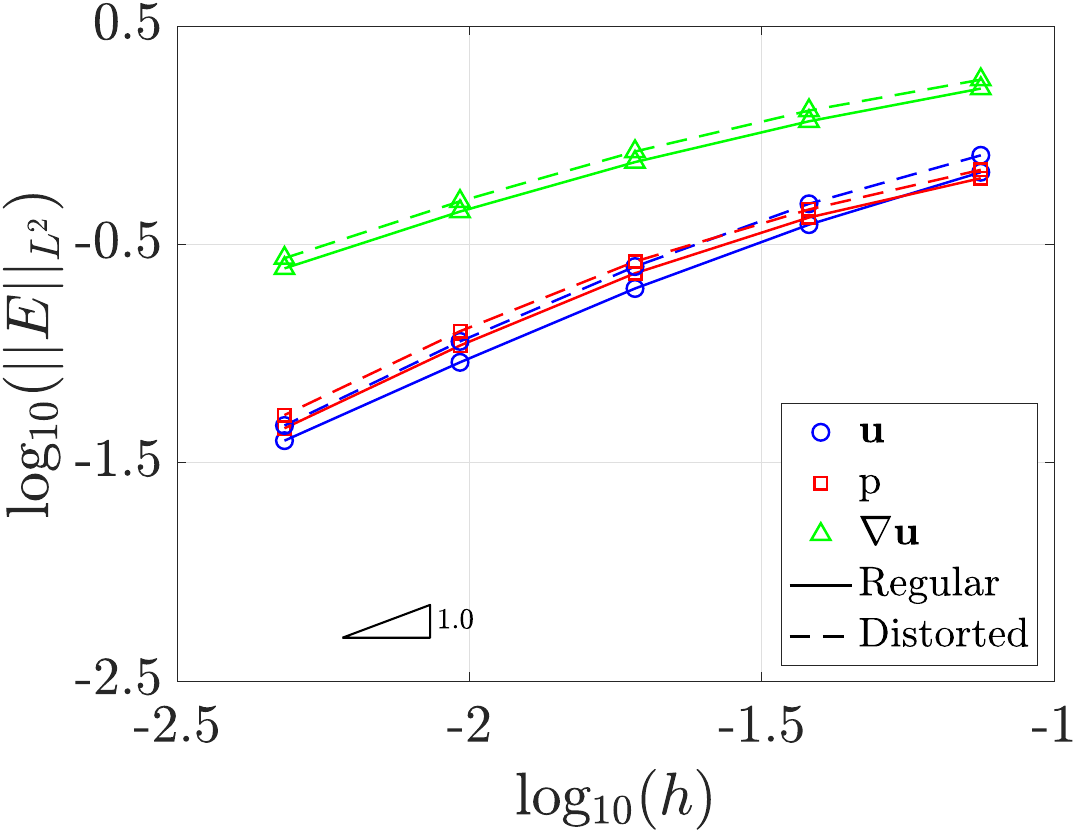}}
	\qquad
	\subfigure[CCFV, $Re {=} 1,000$.]{\includegraphics[width=0.45\textwidth]{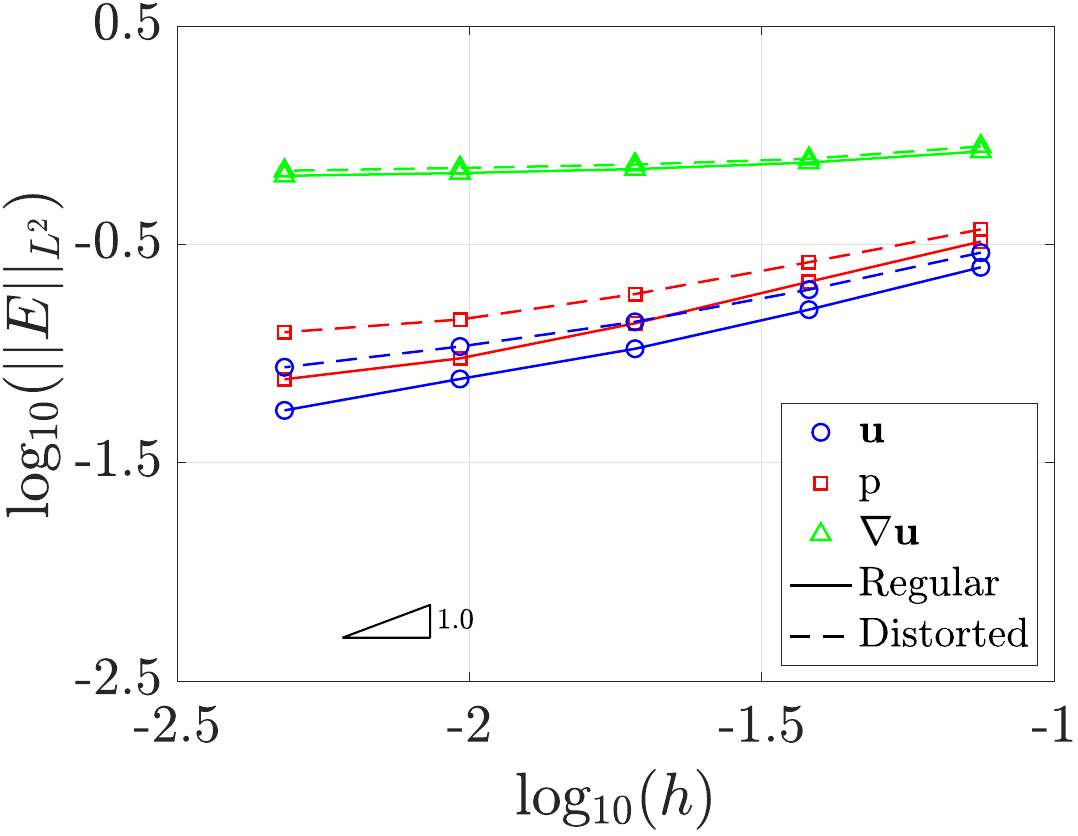}}
	
	\caption{Mesh convergence of the $\eltwo$ errors for the cavity flow.}
	\label{fig:CavTRI}
\end{figure}

The relative error of velocity, pressure, and gradient of velocity is reported for $Re {=} 100$ and $Re {=} 1,000$ in Figure~\ref{fig:CavTRI}.
To compute these errors,  a reference solution obtained using a Taylor-Hood finite element solver on a mesh of $700 {\times} 700$ cells with local refinement near the boundaries is employed.
The errors are computed in a subdomain of $\Omega$ excluding regions $[0,0.05] {\times} [0.95,1]$ and $[0.95,1] {\times} [0.95,1]$ to avoid inconsistencies between the reference solution and the FCFV and CCFV approximations. A detailed discussion on this topic is available in~\cite{vieira2024}.

For both values of the Reynolds number, the OpenFOAM FCFV solver provides first-order accuracy of velocity, pressure, and gradient of velocity, independently of cell distortion. Indeed, the method preserves both the optimal convergence and the accuracy, confirming its robustness to mesh non-orthogonality, skewness, as well as stretching.
On the contrary,  the errors computed using OpenFOAM CCFV scheme stagnate for all tested configurations, with \texttt{simpleFoam} being unable to converge and experiencing increased error when mesh cells are distorted.
This confirms the results of the previously presented experiments, showing the suitability of OpenFOAM FCFV approach to simulate also convection-dominated flows.

\subsection{Oscillating flow in a sweeping jet fluidic oscillator}
\label{sc:Oscillator}

\begin{figure}[!htb]
\centering
\begin{tikzpicture}[scale=0.6]
    \draw [line width = 1.5, line join=round] (-12,1) -- (-8.75,1) -- (-5.75,0.35) -- (-5.75,3.25) -- (-0.5,3.25) -- (-0.5,1.55) -- (0,0.5) -- (0.5,1.55) -- (0.5,5) -- (8,5) -- (8,-5) -- (0.5,-5) -- (0.5,-1.55) -- (0,-0.5) -- (-0.5,-1.55) -- (-0.5,-3.25) -- (-5.75,-3.25) -- (-5.75,-0.35) -- (-8.75,-1) -- (-12,-1) -- cycle;
    \draw [line width = 1.5, line join=round] (-4.75,2.25) -- (-1.5,2.25) -- (-1.5,1.25) -- (-4,1.25) -- (-4.75,0.5) -- cycle;
    \draw [line width = 1.5, line join=round] (-4.75,-2.25) -- (-1.5,-2.25) -- (-1.5,-1.25) -- (-4,-1.25) -- (-4.75,-0.5) -- cycle;
    \draw [line width = 1.5, color = red] (-12,-1) -- node[midway,right]{$\Ga{\text{in}}$} (-12,1);
    \draw [line width = 1.5] (-5.75,-3.25) -- node[midway,left]{$\Ga{\text{w}}$} (-5.75,-0.35);
    \draw [line width = 1.5, color = blue] (8,5) -- node[midway,left]{$\Ga{\text{out}}$} (8,-5);
    \draw[|-|,line width = 0.5] (0.4,0.5) -- node[midway, right]{$D$} (0.4,-0.5);
    \draw[|-|,line width = 0.5] (-5.4,0.35) -- node[midway, right]{$0.7D$} (-5.4,-0.35);
    \draw[|-|,line width = 0.5] (-1.2,1.25) -- node[midway, left]{$2.5D$} (-1.2,-1.25);
    \draw[|-|,line width = 0.5] (1.5,1.55) -- node[midway, right]{$3.1D$} (1.5,-1.55);
    \draw[|-|,line width = 0.5] (3.5,3.25) -- node[midway, right]{$6.5D$} (3.5,-3.25);
    \draw[|-|,line width = 0.5] (-12,-5.3) -- node[midway, below]{$5D$} (-8.75,-5.3);
    \draw[-|,line width = 0.5] (-8.75,-5.3) -- node[midway, below]{$3D$} (-5.75,-5.3);
    \draw[-|,line width = 0.5] (-5.75,-5.3) -- node[midway, below]{$5.25D$} (-0.5,-5.3);
    \draw[-|,line width = 0.5] (-0.5,-5.3) -- node[midway, below]{$D$} (0.5,-5.3);
    \draw[-|,line width = 0.5] (0.5,-5.3) -- node[midway, below]{$36D$} (8,-5.3);
    \draw[|-|,line width = 0.5] (-5.75,4.5) -- node[midway, above]{$D$} (-4.75,4.5);
    \draw[-|,line width = 0.5] (-4.75,4.5) -- node[midway, below]{$0.75D$} (-4,4.5);
    \draw[-|,line width = 0.5] (-4,4.5) -- node[midway, above]{$2.5D$} (-1.5,4.5);
    \draw[-|,line width = 0.5] (-1.5,4.5) -- node[midway, above]{$D$} (-0.5,4.5);
    \draw[|-|,line width = 0.5] (-12.3,1) -- node[midway, left]{$2D$} (-12.3,-1);
    \draw[|-|,line width = 0.5] (8.3,5) -- node[midway, right]{$40D$} (8.3,-5);
    \draw[|-|,line width = 0.5] (-6.6,3.25) -- node[midway, left]{$D$} (-6.6,2.25);
    \draw[|-|,line width = 0.5] (-5.25,-4) -- node[midway, below]{$\Lf$} (-1,-4);
\end{tikzpicture}
\caption{Computational domain and boundary conditions for the fluidic oscillator.}
\label{fig:FOgeom}
\end{figure}
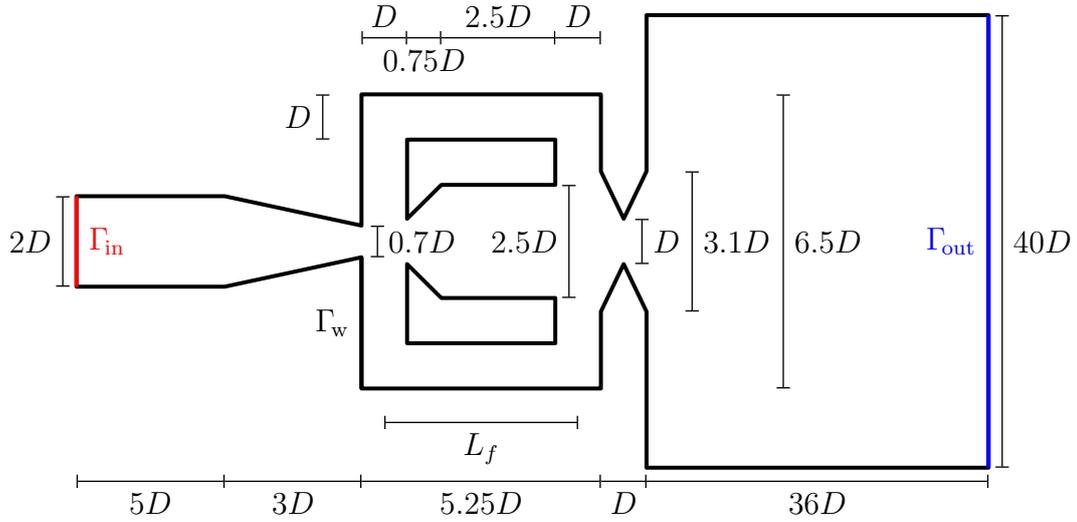

The last example consists of a fluidic oscillator (see Figure~\ref{fig:FOgeom}),  converting a steady input jet into an oscillating one without any moving part.
A uniform, horizontal velocity profile is imposed on the inlet boundary $\Ga{\text{in}}$, homogeneous Neumann boundary conditions are applied at the outlet boundary $\Ga{\text{out}}$, and no-slip conditions are enforced on all remaining walls denoted by $\Ga{\text{w}}$.
The diameter $D{=}1$ and average velocity $\Uo{=}1$ at the outlet nozzle are selected as characteristic length and characteristic velocity of the problem, the corresponding value of the Reynolds number being $Re {=} 5,000$.

\begin{figure}[!htb]
	\centering
	\subfigure[Mesh.]{\includegraphics[width=0.44\textwidth]{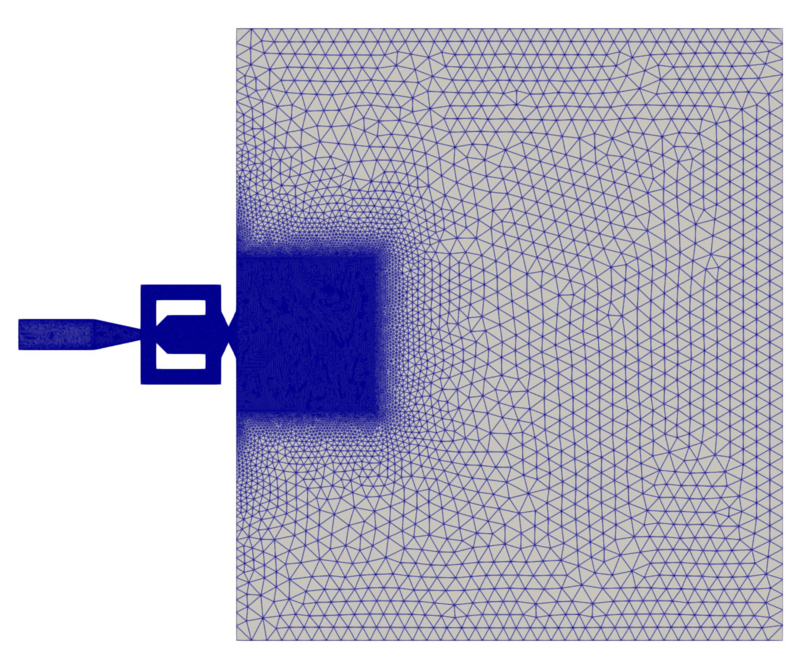}}
        \qquad
	\subfigure[Zoom inside the oscillator.\label{fig:FOmeshZoom}]{\includegraphics[width=0.42\textwidth]{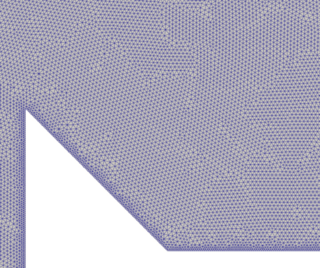}}
 
	\caption{Computational mesh for the fluidic oscillator.}
	\label{fig:FOmesh}
\end{figure}

Note that the complex geometric features of this device make particularly challenging the generation of a structured, orthogonal mesh of quadrilateral cells representing the optimal environment for OpenFOAM CCFV simulation. 
On the contrary,  unstructured meshes of simplices can be easily obtained using automatic mesh generation software.
The mesh, designed with the software \texttt{GiD},  was constructed using a Delaunay algorithm. 
The local mesh size is inspired by~\cite{seo2018}, where a second-order finite difference method was used on a uniform mesh with characteristic cell size $D/24$.
In this work, a mesh size of $D/48$ is employed inside the oscillator, with the smallest cells of size $D/96$ being located in the vicinity of the boundaries.
A cell size of $D/24$ is set in the convergent nozzle at the inlet and in the refined square region after the outlet nozzle.
Finally,  mesh size of $D/12$ and $D$ are imposed on the remaining parts of inlet and outlet regions, respectively. 
The resulting mesh consists of approximately $320,000$ cells and is displayed in Figure~\ref{fig:FOmesh}.

\begin{figure}[!htb]
	\centering
	\subfigure[$\displaystyle\frac{1}{6}T$.]{\includegraphics[width=0.3\textwidth]{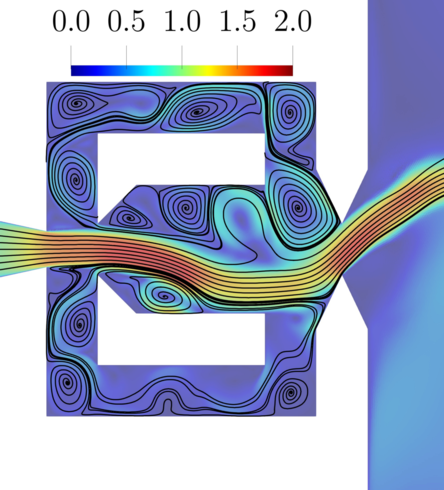}}
        \hfill
	\subfigure[$\displaystyle\frac{1}{3}T$.]{\includegraphics[width=0.3\textwidth]{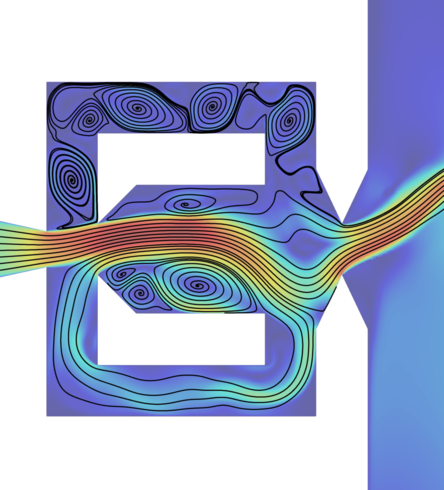}}
        \hfill
	\subfigure[$\displaystyle\frac{1}{2}T$.]{\includegraphics[width=0.3\textwidth]{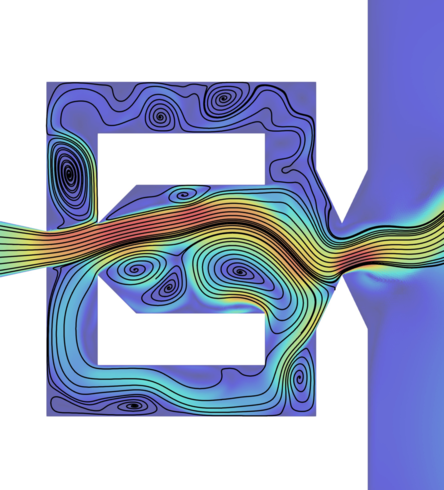}}
 
	\subfigure[$\displaystyle\frac{2}{3}T$.]{\includegraphics[width=0.3\textwidth]{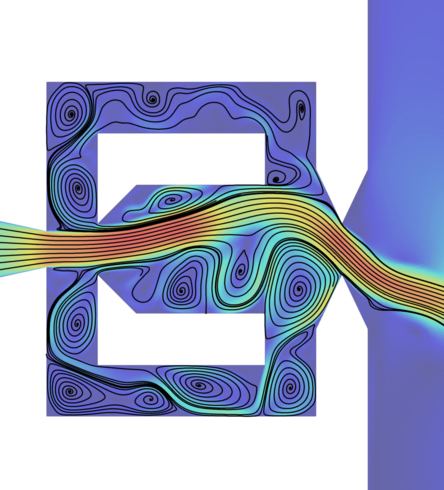}}
        \hfill
	\subfigure[$\displaystyle\frac{5}{6}T$.]{\includegraphics[width=0.3\textwidth]{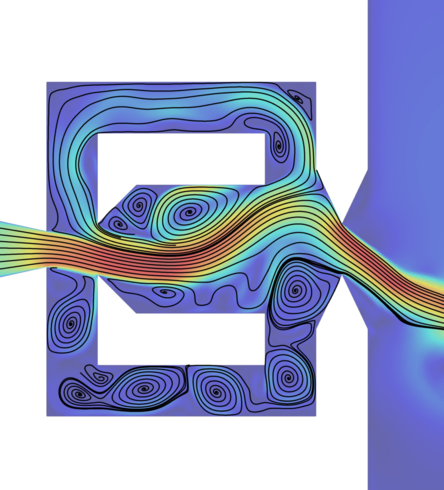}}
        \hfill
	\subfigure[$T$.]{\includegraphics[width=0.3\textwidth]{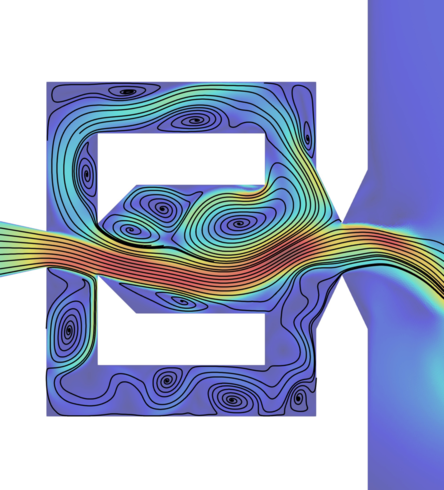}}
	\caption{Magnitude of velocity and streamlines in the fluidic oscillator computed using FCFV.}
	\label{fig:FOflow}
\end{figure}

The FCFV simulation is executed for $1,000$ non-dimensional time units with a BDF2 time integrator and constant $\Delta t {=} 3 {\times} 10^{-2}$, corresponding to a Courant number of approximately $10$.
The solutions are sampled every $0.33$ time units for both OpenFOAM FCFV solver and \texttt{pimpleFoam}.
Figure~\ref{fig:FOflow} reports the magnitude of the velocity field and the streamlines computed using OpenFOAM FCFV solver at different time instants.
The results show a recurring oscillatory flow pattern with a characteristic period $T$.  The primary flow, characterised by the largest volume of fluid, oscillates in the mixing chamber between the top and bottom walls.  Concurrently, recirculation regions form on the opposite side of the primary flow and secondary flows, with smaller volume of fluid, are observed in the feedback channels: depending on the direction of the primary flow,  the flow in either one channel or the other is favoured, while additional vortices are formed in the remaining feedback channel.

\begin{figure}[!htb]
    \centering
    \subfigure[FCFV: u$_y$ and $\Delta$p.]{\includegraphics[width=0.55\textwidth]{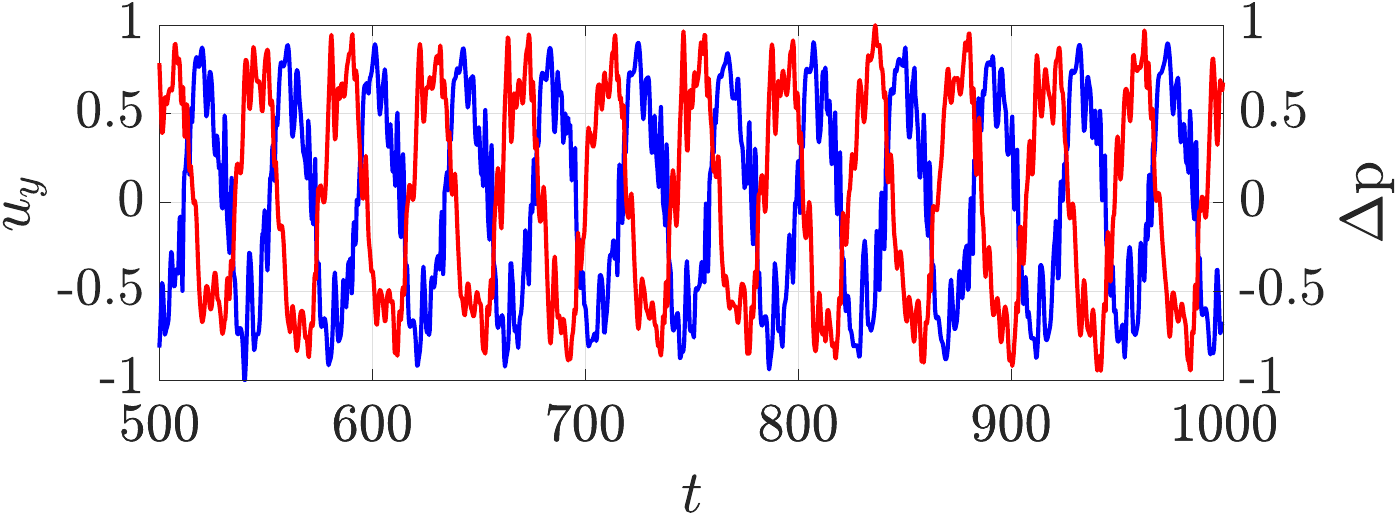}}
    \quad
    \subfigure[FCFV: frequency.]{\includegraphics[width=0.4\textwidth]{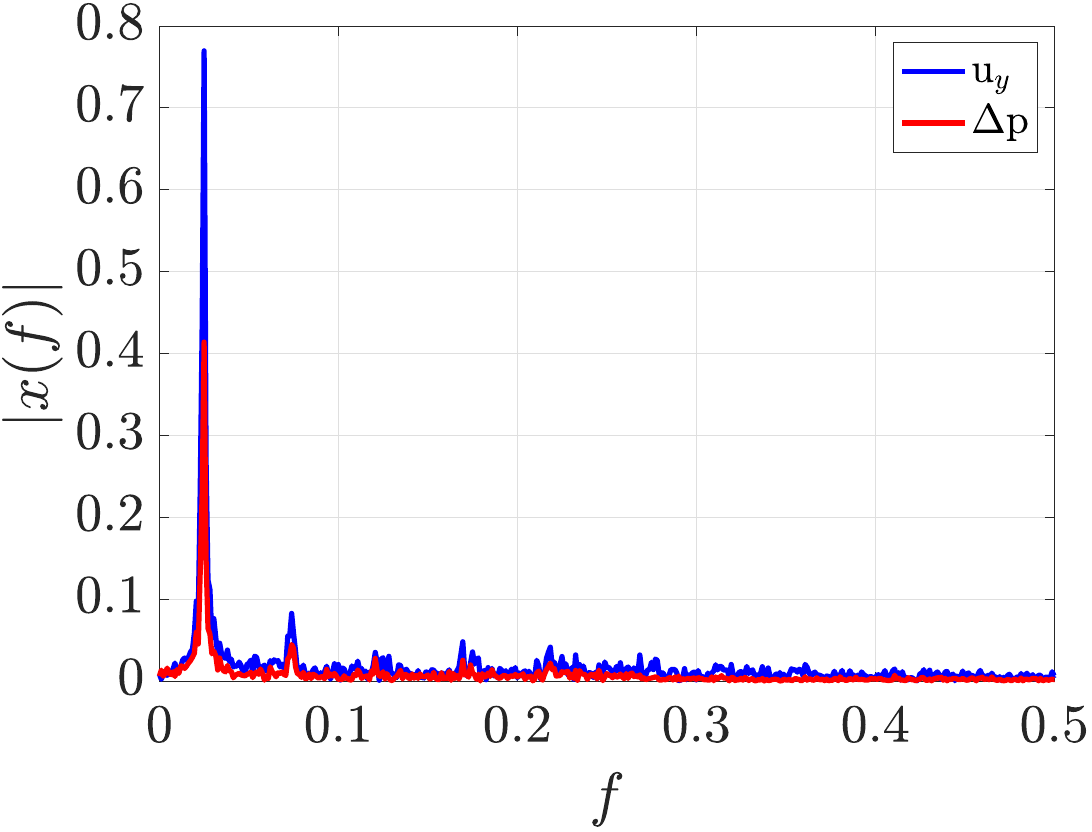}\label{fig:FOresultsB}}

    \subfigure[CCFV: u$_y$ and $\Delta$p.]{\includegraphics[width=0.55\textwidth]{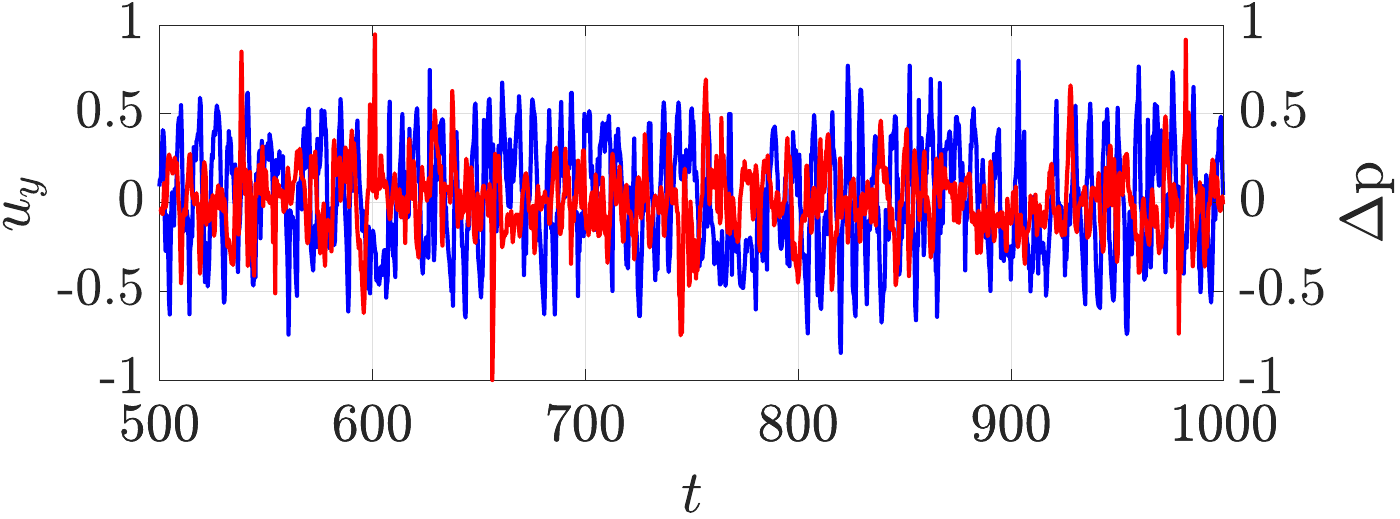}}
    \quad
    \subfigure[CCFV: frequency.]{\includegraphics[width=0.4\textwidth]{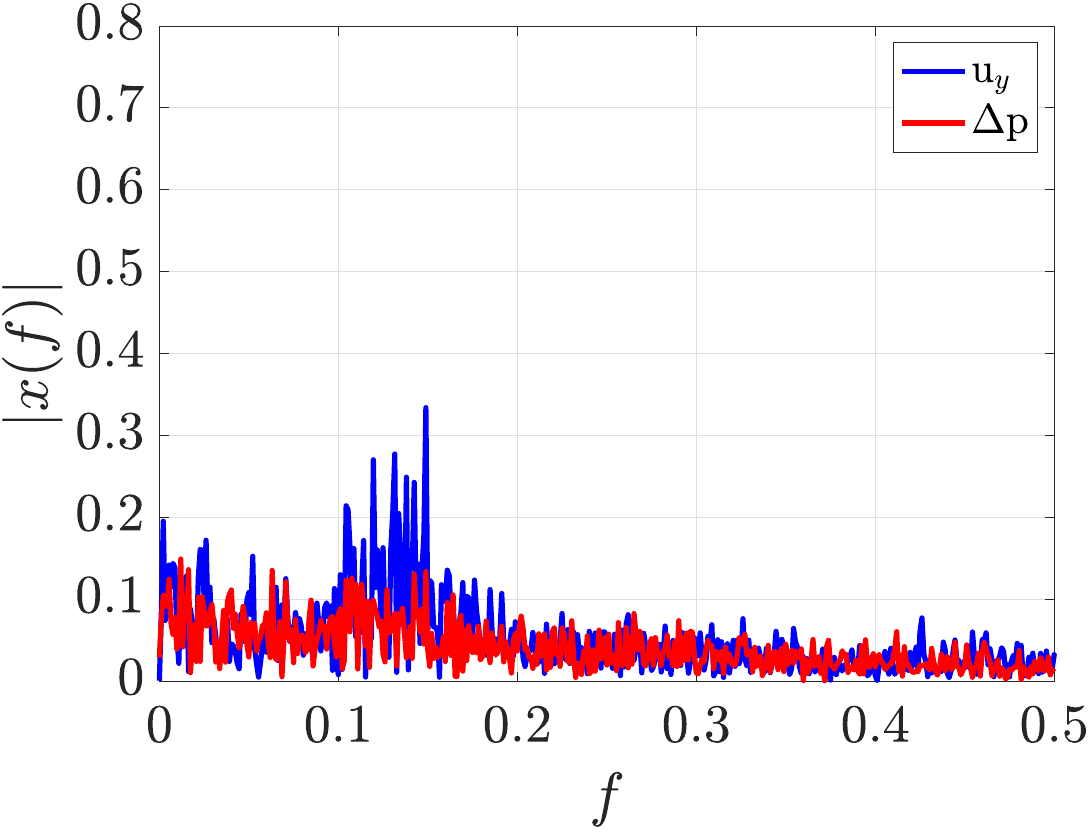}\label{fig:FOresultsD}}
    
    \caption{Simulation of the fluidic oscillator. Left: temporal evolution of the non-dimensional velocity and pressure difference.  Right: Frequencies spectra.}
    \label{fig:FOresults}
\end{figure}

To quantitatively analyse this problem, the frequency $f$ of the oscillations is measured, either using the vertical component of the velocity at the middle of the outlet nozzle or by means of the pressure difference at the inlets of the feedback channels. The corresponding non-dimensional Strouhal number is defined as $St {=} fD/\Uo$.
Figure~\ref{fig:FOresults} presents a comparison of FCFV and CCFV outputs.
The FCFV results display a periodic behaviour of both vertical velocity and pressure difference, with a clear dominant frequency as visible in Figure~\ref{fig:FOresultsB}. In addition, higher-frequency fluctuations are identified, consistently with experimental studies, see, e.g., \cite{woszidlo2015}.
On the contrary,  the solution obtained using \texttt{pimpleFoam} exhibits a highly noisy behaviour, failing to identify any dominant frequency as displayed in Figure~\ref{fig:FOresultsD}.
It is worth noticing that this limitation of \texttt{pimpleFoam} is not associated with the use of an unstructured simplicial mesh. Numerical experiments, not reported here for brevity,  were performed using a mesh of quadrilateral cells designed with \texttt{snappyHexMesh} and no clear improvement was observed in the approximation of the velocity and pressure fields, nor in the computation of the frequency of the problem.
In both setups,  mesh non-orthogonality is limited with an average non-orthogonality angle of $5$ and $2$ degrees for triangular and quadrilateral meshes, respectively. Nonetheless,  face skewness appears to be particularly relevant, especially in regions such as the one displayed in Figure~\ref{fig:FOmeshZoom}: despite the number of skewed cells is limited and they are mainly concentrated in the vicinity of the boundaries, their presence is sufficient to prevent \texttt{pimpleFoam} to accurately compute the solution.
This confirms the impact on CCFV approximations of the error induced by mesh non-conjunctionality, which does not significantly affect the discussed FCFV framework.

\begin{figure}[!htb]
        \centering
        \includegraphics[width=0.5\textwidth]{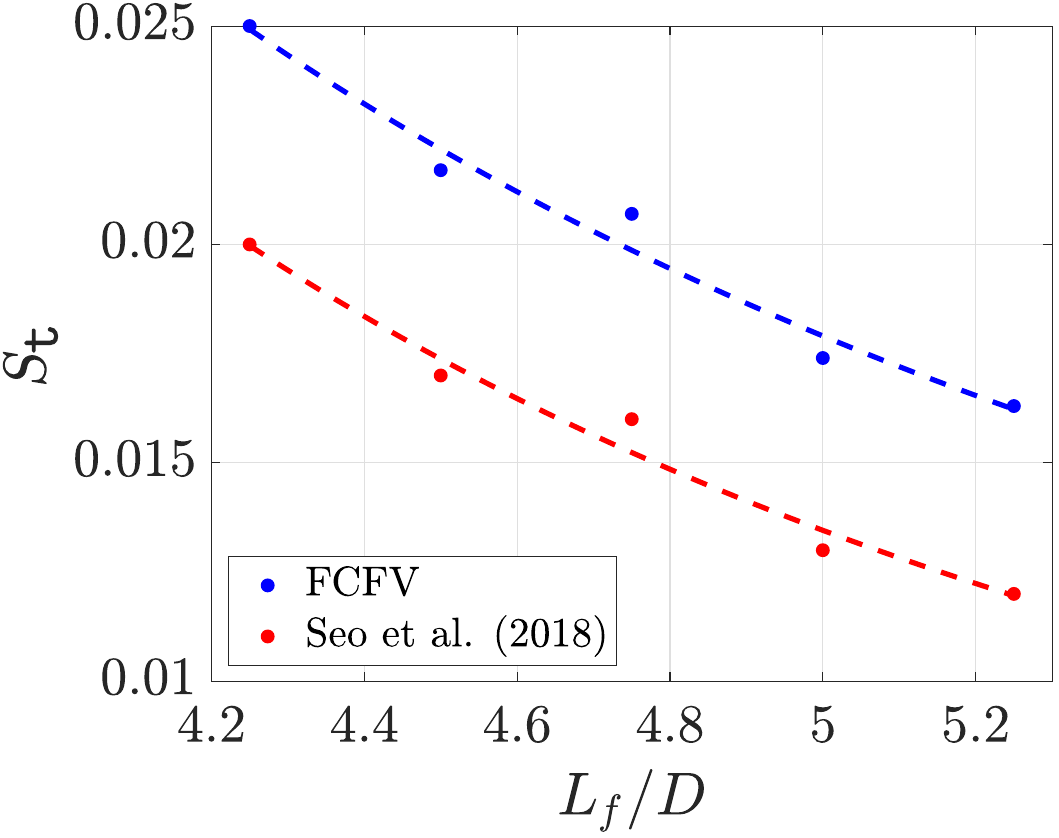}
        \caption{Strouhal number for different configurations.}
        \label{fig:FOstr}
\end{figure}

Finally, following~\cite{seo2018,tajik2021}, the effect of the mixing chamber length $\Lf$, see Figure~\ref{fig:FOgeom}, on the oscillation frequency is studied.
Five geometric configurations are tested, with $\Lf/D$ ranging from $4.25$ to $5.25$.
The results obtained using the OpenFOAM FCFV solver are reported in Table~\ref{tab:FreqOsc}, with the corresponding values from~\cite{seo2018}. Note that the reference does not disclose the exact geometric schematics of the device due to patent restrictions, thus only the qualitative behaviour of the Strouhal number can be compared.
Indeed,  Figure~\ref{fig:FOstr} displays the Strouhal number computed for different values of the parameter $\Lf/D$ and the results reported by Seo \etal in~\cite{seo2018}, along with their corresponding best-fit power laws.
The results show excellent agreement with the reference, confirming the capability of the OpenFOAM FCFV scheme to identify the trend discussed in previously published works in the literature, while an almost constant deviation is observed between computed and reference values due to the uncertain geometric configurations.
 
\begin{table}[!htb]
   \centering
        \begin{tabular}{c c c}
            \hline
            \multirow{2}{*}{$\displaystyle\frac{\Lf}{D}$} & \multicolumn{2}{c}{$St$} \\
		 & FCFV & Ref. \cite{seo2018}  \\
		\hline
            4.25 & 0.0250 & 0.020 \\
            4.50 & 0.0217 & 0.017 \\
            4.75 & 0.0207 & 0.016 \\
            5.00 & 0.0174 & 0.013 \\
            5.25 & 0.0163 & 0.012 \\
            \hline
	\end{tabular}
\caption{Dependence of $St$ on $\Lf/D$.}
\label{tab:FreqOsc}
\end{table}

\section{Conclusions}
\label{sc:Conclusion}

An OpenFOAM solver robust to mesh-induced errors was presented employing the face-centred FV paradigm.
The method avoids gradient reconstruction,  reducing its sensitivity to well-known issues of cell-centred FV schemes on non-orthogonal, stretched, and skewed meshes.
The integration of the FCFV solver in the open-source library OpenFOAM allows to seamlessly employ this methodology, while reducing the complexity of generating meshes suitable for CFD computations.
Numerical benchmarks showed that FCFV preserves, also for convection-dominated problems,  accuracy, stability,  robustness, and optimal convergence in the presence of cell distortion, where the quality of CCFV approximations significantly degrades.

\subsection*{Acknowledgments}
Generalitat de Catalunya (PhD scholarship of DC; 2021-SGR-01049 to MG and AH; Serra H\'unter Programme for MG). Spanish Ministry of Science, Innovation and Universities and Spanish State Research Agency MICIU/AEI/10.13039/501100011033 (PID2020-113463RB-C33 to MG; PID2020-113463RB-C32 to AH; CEX2018-000797-S to MG and AH).


\end{document}